**"WM"-Shaped Growth of GaN on Patterned Sapphire Substrates**

Lai Wang[*], Xiao Meng, Di Yang, Zhibiao Hao, Yi Luo, Changzheng Sun, Yanjun Han, Bing Xiong, Jian Wang and Hongtao Li

*State Key Laboratory on Integrated Optoelectronics / Tsinghua National Laboratory for Information Science and Technology, Department of Electronic Engineering, Tsinghua University, Beijing 100084, P. R. China*

[*] Corresponding author. Tel: +86-10-62798240; Fax: +86-10-62784900

*E-mail address:* wanglai@tsinghua.edu.cn (L. Wang)

**Abstract**

In metal organic vapor phase epitaxy of GaN, the growth mode is sensitive to reactor temperature. In this study, V-pit-shaped GaN has been grown on normal c-plane cone-patterned sapphire substrate by decreasing the growth temperature of high-temperature-GaN to around 950 °C, which leads to the 3-dimensional growth of GaN. The so-called "WM" well describes the shape that the bottom of GaN V-pit is just right over the top of sapphire cone, and the regular arrangement of V-pits follows the patterns of sapphire substrate strictly. Two types of semipolar facets $(1\bar{1}01)$ and $(11\bar{2}2)$ expose on sidewalls of V-pits. Furthermore, by raising the growth temperature to 1000 °C, the growth mode of GaN can be transferred to 2-demonsional growth. Accordingly, the size of V-pits becomes smaller and the area of c-plane GaN becomes larger, while the total thickness of GaN keeps almost unchanged during this process. As long as the 2-demonsional growth lasts, the V-pits will disappear and only flat c-plane GaN remains. This means the area ratio of c-plane and semipolar plane GaN can be controlled by the duration time of 2-demonsional growth.

1. Introduction

Sapphire is the most widely used substrate for III-nitride semiconductors growth. Even though there is a large lattice mismatch of about 16% between GaN and sapphire, "two-step" growth method, invented by Amano and Akasaki first and further developed by Nakamura later,[1, 2] has been proven efficient to improve the crystal quality of GaN grown on sapphire substrate. Tadatomo et al. in 2001 introduced patterned sapphire substrate (PSS) to growth GaN-based light-emitting diodes (LED) and achieved the enhanced light extraction efficiency,[3] since total reflection of light at the interface between GaN and sapphire is destroyed. Then, it was also found that PSS can reduce the thread dislocation density in GaN by controlling the growth mode of GaN and making dislocation turn to parallel to substrate.[4,5] Up to now, PSS has been a standard process for horizontal LED chip fabrication in industry. The growth technique of high quality and flat GaN on PSS has been developed maturely.

On the other hand, some researchers also tried to deposit patterned $SiO_2$ or $Si_xN$ mask on flat GaN and proceed regrowth of GaN on it, which is also called selective area growth (SAG).[6-8] The GaN growth mode could be controlled by growth parameters, such as reactor pressure, temperature, V/III ratio, etc. Through implementing epitaxial lateral overgrowth (ELOG) of GaN on mask, dislocation density in follow-up GaN could be reduced further.[9, 10] In addition, by using 3-dimensional (3D) growth mode in SAG process, it can make GaN expose its semipolar planes.[11, 12] This growth method of semipolar GaN is competitive since it could avoid stacking faults in GaN or extremely expensive semipolar GaN substrate. However, the fabrication of mask and regrowth of GaN increase the complexity of process.

In this paper, we exhibit a method to grow semipolar GaN on PSS directly. By moderately decreasing the growth temperature of GaN in metal organic vapor phase epitaxy (MOVPE) reactor, V-pit-shaped GaN can be obtained on normal PSS due to 3D growth mode. The V-pits arrange regularly and follow the patterns of substrate strictly, wherein the bottom of GaN V-pit is just over the top of sapphire cone, like "WM" letters on the top and bottom. The exposed sidewalls of V-pits, $(1\bar{1}01)$ and $(11\bar{2}2)$ semipolar facets, dominate the surface, while the c-plane GaN only exists among the space of close-packed V-pits. When the growth temperature is raised to normal value, the growth mode can be transferred to 2-demonsional (2D) growth, leading to size diminishment of V-pits. As long as the 2D growth lasts, the V-pits will become smaller and disappear finally.

2. Experiments

MOVPE growth is carried out in an AIXTRON 2000HT reactor. Standard commercialized 2-inch c-plane PSS are used, wherein the patterns are hexagonal close-packed cones with 3-μm period, 2.7-μm bottom diameter and 1.7-μm height. GaN is grown on PSS by "two-step" method. First, the PSS is cleaned under 1100 °C in $H_2$ ambient for 10 minutes. Then, a 30-nm GaN buffer layer is deposited under 520 °C. Next, high-temperature GaN is grown under 950 °C for 1 hour to form V-pits on surface. This temperature is slightly lower than normal value above 1000 °C, leading to 3D growth mode. For some samples, the growth process is finished here, while for others, subsequent growth under 1000 °C continues lasting for 5 to 30 minutes. In this case, the growth mode will transfer to 2D growth, resulting in diminishment of V-pits. After growth, the surface and cross-sectional morphologies of samples are observed by optical microscopy and scanning electron microscopy (SEM).

3. Results and Discussion

The schematic diagram of "WM"-shaped growth of GaN is plotted in Fig. 1(a). It should be noticed that the side surface of cone on PSS prepared by inductively coupled plasma (ICP) etching has certain radian. Thus, there is no large area stable crystal face existing on the side surface of cone. When GaN is grown on PSS, nucleation can be only implemented on c-plane surface of sapphire, resulting in island-like GaN distributing in the gap between adjacent cones. In normal growth process of GaN, as the temperature is above 1000 °C, 2D growth of GaN will promote the lateral conjoining between these GaN islands and form a flat and smooth c-plane surface of GaN finally. However, in "WM"-shaped growth process, since the growth temperature is 950 °C. Under this condition, the growth rate of c-plane will be much higher than that of other semipolar planes. As growth continues, c-plane will become smaller gradually, while V-pits with semipolar facets inside will be formed at the same time. Figure 1(b) shows the SEM photo of GaN grown under 950 °C for 15 minutes on PSS. It is clear that GaN islands are grown on c-plane sapphire regions while no GaN nucleation occurs on the side surface of the cones. Owing to the 3D growth, abrupt semipolar facets of GaN have been formed.

The surface morphology photo of the sample after 3D growth for 1 hour is shown in Fig. 2 (a). Regular hexagonal V-pits arranging according to the patterns of substrate are clearly observed. The V-pits occupy the main area of surface, while the triangular c-plane GaN only exists among the space of adjacent V-pits. The details of V-pits are further characterized by SEM, as shown in Fig.2 (b)~(e). In SEM photos, the semipolar facets inside V-pits can be divided to two types, bright regions and dark regions, and the former dominates the sidewalls. According to their orientation and the title angle from Fig. 2 (e), it is easy to judge that they are $(1\bar{1}01)$ and $(11\bar{2}2)$ facets, respectively. These two types of semipolar facets are typical ones, which were also reported in samples grown by SAG.[11-16] In Fig. 2 (e), it is also found that the bottom of V-pit is just over the top of sapphire cone, which further confirms the principle of "WM"-shaped growth explained in Fig. 1. The total thickness of GaN is about 4 μm. For comparison, we also try to grow GaN under 950 °C on planar sapphire substrate. The dense smaller V-pits defects can also be observed on surface, as shown in Fig. 2(f). This phenomenon implies that the 3D growth mode is driven by the lower temperature, even if no PSS is used. Generally, a thread dislocation is believed existing beneath a V-pit. However, the V-pits formed in this case distribute randomly and their sizes are also nonuniform, which means their controllability is not as high as those grown on PSS.

Based on the V-pits obtained in Fig. 2 (a) and (b), 2D growth under 1000 °C is carried out subsequently to diminish the size of V-pits. This temperature is still lower than the normal value of about 1020-1050 °C for high quality GaN growth. Higher growth temperature would accelerate the 2D growth process. However, in order to ensure the controllability, we choose 1000 °C in our experiment. Figure 3 (a)~(e) show the surface morphologies of samples with different 2D growth time of duration. For 5-minutes and 10-minutes growth, the V-pits are diminished gradually, while the area of c-plane GaN is expanded accordingly, as shown in Fig. 3(a) and (b). It is also can be seen that the uniformity of V-pits is still kept well. Compared with Fig. 2(d), it seems that the area of dark regions, corresponding to $(11\bar{2}2)$ facets, increase slightly. For the sample after 15-minutes 2D growth, the V-pits are diminished further, but their uniformity starts to become worse, and their shapes also get a little irregular. When 2D growth lasts for 20 minutes, the distribution uniformity and shape irregularity of V-pits deteriorate further. Some V-pits have been filled very small and are difficult to be seen under optical microscope. In the sample for 30-minutes 2D growth, most V-pits are fully filled and disappear, while some sparse small V-pits can be still observed in SEM photo. It should be point out that the thickness of GaN observed from cross-sectional TEM photos (not shown) for samples in Fig. 3(a)~(d) are all around 4 μm, the same with the sample in Fig. 2(e), while that for the sample in Fig. 3(e) is about 4.6 μm. This result indicates that the during the initial stage of 2D growth, the adatoms mainly migrate to V-pits and incorporate into lattice there. This implies that under this growth condition, the migration length of adatoms is longer than the distance between adjacent V-pits. However, when most V-pits are filled, the distance between two V-pits becomes longer than the migration length, and hence the 2D growth starts to increase the thickness of GaN film. It could be expected that when 2D growth lasts continuously, the surface of GaN will become flatter and smoother without any V-pit.

The present experiments are carried out based on standard commercialized PSS. Through defining the different periods and arrangements of patterns on sapphire substrate, the V-pits with various sizes and arrangements can be prepared by 3D growth. And by controlling the time duration of 2D growth, the duty ratio of V-pits on surface can also be well controlled. For example, if the nano-PSS is used, wherein the period of patterns is only several hundreds of nanometers, the size of V-pit can be further reduced. Therefore, this "WM"-shaped growth method is expected for many novel optoelectronic or photonic devices applications, such as semipolar plane light-emitting devices, photonic crystals, etc.

4. Conclusions

GaN with V-pits on surface has been grown on PSS by decreasing the growth temperature to around 950 °C in MOVPE. The V-pits arranged regularly and follow the patterns of substrate strictly. The bottom of V-pit is just over the top of sapphire cone, which indicates the V-pits are formed as a result of 3D growth under 950 °C. The growth rate of c-plane GaN is much faster, leading to exposure of semipolar planes. The sidewalls of V-pits are $(1\bar{1}01)$ and $(11\bar{2}2)$ facets. When increasing the growth temperature to 1000 °C subsequently, the growth mode will transfer to 2D growth and result in diminishing of V-pits. Before V-pits disappear, the total thickness of GaN will kept unchanged. As 2D growth lasts, the V-pits will be fully filled and disappear, while the thickness of GaN starts to increase.

Acknowledgement

The authors are thankful for the supports of National Key Research and Development Plan (Grant No. 2016YFB0400102), National Basic Research Program of China (Grant Nos. 2012CB3155605, 2013CB632804, 2014CB340002 and 2015CB351900), the National Natural Science Foundation of China (Grant Nos. 61574082, 61210014, 61321004, 61307024, and 51561165012), the High Technology Research and Development Program of China (Grant No. 2015AA017101), Tsinghua University Initiative Scientific Research Program (Grant No. 2013023Z09N, 2015THZ02-3), the Open Fund of the State Key Laboratory on Integrated Optoelectronics (Grant No. IOSKL2015KF10), the CAEP Microsystem and THz Science and Technology Foundation (Grant No. CAEPMT201505) and S&T Challenging Project.

References

[1] H. Amano, N. Sawaki, I. Akasaki, and Y. Toyoda, Metalorganic vapor phase epitaxial growth of a high quality GaN film using an AlN buffer layer, Applied Physics Letters 48 (1986) 353.

[2] S. Nakamura, GaN Growth using GaN buffer layer, Japanese Journal of Applied Physics, 30 (1991) L1705.

[3] K. Tadatomo, H. Okagawa, Y. Ohuchi, T. Tsunekawa, Y. Imada, M. Kato and Tsunemasa Taguchi, High output power InGaN ultraviolet light-emitting diodes fabricated on patterned substrates using metalorganic vapor phase epitaxy, Japanese Journal of Applied Physics 40 (2001) 6B

[4] J.-C. Song, S.-H Lee, I.-H. Lee, K.-W. Seol, S. Kannappan, and C.-R. Lee, Characteristics comparison between GaN epilayers grown on patterned and unpatterned sapphire substrate (0001), Journal of Crystal Growth 308 (2007) 321–324.

[5] H.-Y. Shin, S. K. Kwon, Y. I. Chang, M. J. Cho, and K. H. Park, Reducing dislocation density in GaN films using a cone-shaped patterned sapphire substrate, Journal of Crystal Growth 311 (2009) 4167–4170.

[6] Y. Kato, S. Kitamura, K. Hiramatsu, and N. Sawaki, Selective growth of wurtzite GaN and $Al_xGa_{1-x}N$ on GaN/sapphire substrates by metalorganic vapor phase epitaxy, Journal of Crystal Growth 144 (1994) 133-140.

[7] D. Kapolnek, S. Keller, R. Vetury, R. D. Underwood, P. Kozodoy, S. P. Den Baars, and U. K. Mishra, Anisotropic epitaxial lateral growth in GaN selective area epitaxy, Appl. Phys. Lett. 71 (1997) 1204.

[8] K. Hiramatsu, K. Nishiyama, A. Motogaito, H. Miyake, Y. Iyechika, and T. Maeda, Recent progress in selective area growth and epitaxial lateral overgrowth of III-nitrides: effects of reactor pressure in MOVPE growth, Physica Status Solidi (a) 176 (1999) 535–543.

[9] S. Nakamura, M. Senoh, S. Nagahama, N. Iwasa, T. Yamada, T. Matsushita, H. Kiyoku, Y. Sugimoto, T. Kozaki, H. Umemoto, M. Sano, and K. Chocho, InGaN/GaN/AlGaN-based laser diodes with modulation-doped strained-layer superlattices grown on an epitaxially laterally overgrown GaN substrate, Applied Physics Letters, 72 (1998) 211.

[10] J. Mei, S. Srinivasan, R. Liu, F. A. Ponce, Y. Narukawa, and T. Mukai, Prismatic stacking faults in epitaxially laterally overgrown GaN, Appl. Phys. Lett. 88 (2006) 141912.

[11] T. Wunderer1, M. Feneberg, F. Lipski, J. Wang, R. A. R. Leute, S. Schwaiger, K. Thonke, A. Chuvilin, U. Kaiser, S. Metzner, F. Bertram, J. Christen, G. J. Beirne, M. Jetter, P. Michler, L. Schade, C. Vierheilig, U. T. Schwarz, A. D. Dräger, A. Hangleiter, and F. Scholz, Three-dimensional GaN for semipolar light emitters, Physica Status Solidi (b) 248 (2011) 549-560.

[12] F. Scholz, Semipolar GaN grown on foreign substrates: a review, Semicond. Sci. Technol. 27 (2012) 024002.

[13] T. Wunderer, F. Lipski, J. Hertkorn, S. Schwaiger, and F. Scholz, Fabrication of 3D InGaN/GaN structures providing semipolar GaN planes for efficient green light emission, Phys. Status Solidi C 6 (2009) S490–S493.

[14] T. Wunderer, F. Lipski, S. Schwaiger, J. Hertkorn, M. Wiedenmann1, M. Feneberg1, K. Thonke1, and F. Scholz, Properties of blue and green InGaN/GaN quantum well emission on structured semipolar surfaces, Japanese Journal of Applied Physics 48 (2009) 060201.

[15] T. Wunderer1, J. Wang, F. Lipski, S. Schwaiger, A. Chuvilin, U. Kaiser, S. Metzner, F. Bertram, J. Christen, S. S. Shirokov, A. E. Yunovich, and F. Scholz, Semipolar GaInN/GaN lightemitting diodes grown on honeycomb patterned substrates, Phys. Status Solidi C 7 (2010) 2140–2143.

[16] S. Metzner, F. Bertram, C. Karbaum, T. Hempel, T. Wunderer, S. Schwaiger, F. Lipski, F. Scholz, C. Wächter, M. Jetter, P. Michler, and J. Christen, Spectrally and time-resolved cathodoluminescence microscopy of semipolar InGaN SQW on (1122) and (1011) pyramid facets, Phys. Status Solidi B 248 (2011) 632–637.

Figure captions

Fig. 1 (a) Schematic diagram of "WM"-shaped growth of GaN on PSS; (b) Scanning electron microscopy photo of high-temperature GaN after 15-minutes growth. The bright is GaN, which is only grown on c-plane sapphire substrate.

Fig. 2 Surface morphology of V-pits on GaN. (a) Optical microscopy photo; (b) SEM photos under low (a) and high (b) magnification; (c) Details in a V-pit; (d) Cross-sectional SEM photo of a V-pit on top of a sapphire cone. (e) V-pits on GaN surface grown on flat sapphire substrate due to a low growth temperature.

Fig. 3 Optical microscopy photos and SEM photos of samples: 3D growth for 1hour + 2D growth for 5 minutes (a), 10 minutes (b), 15 minutes (c), 20 minutes (d), and 30minutes (e).

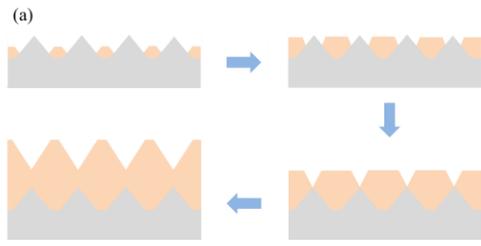 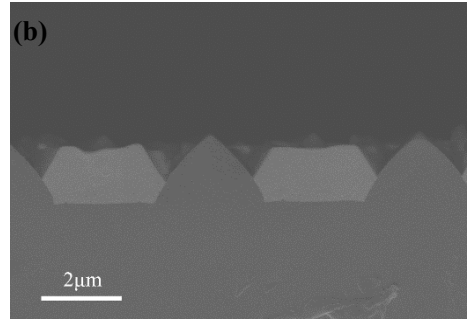

Fig.1 L. Wang *et al.*

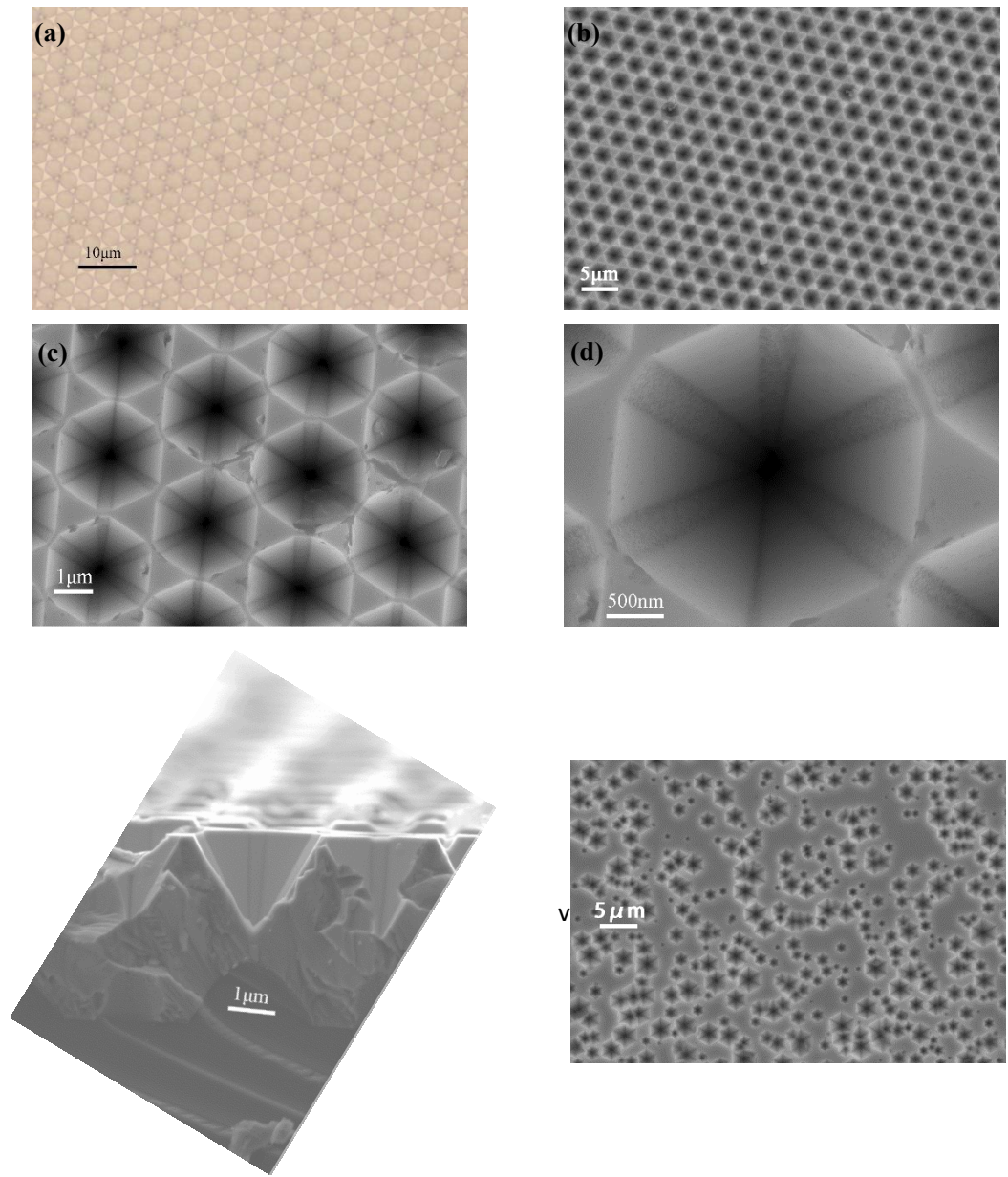

Fig.2 L. Wang *et al.*

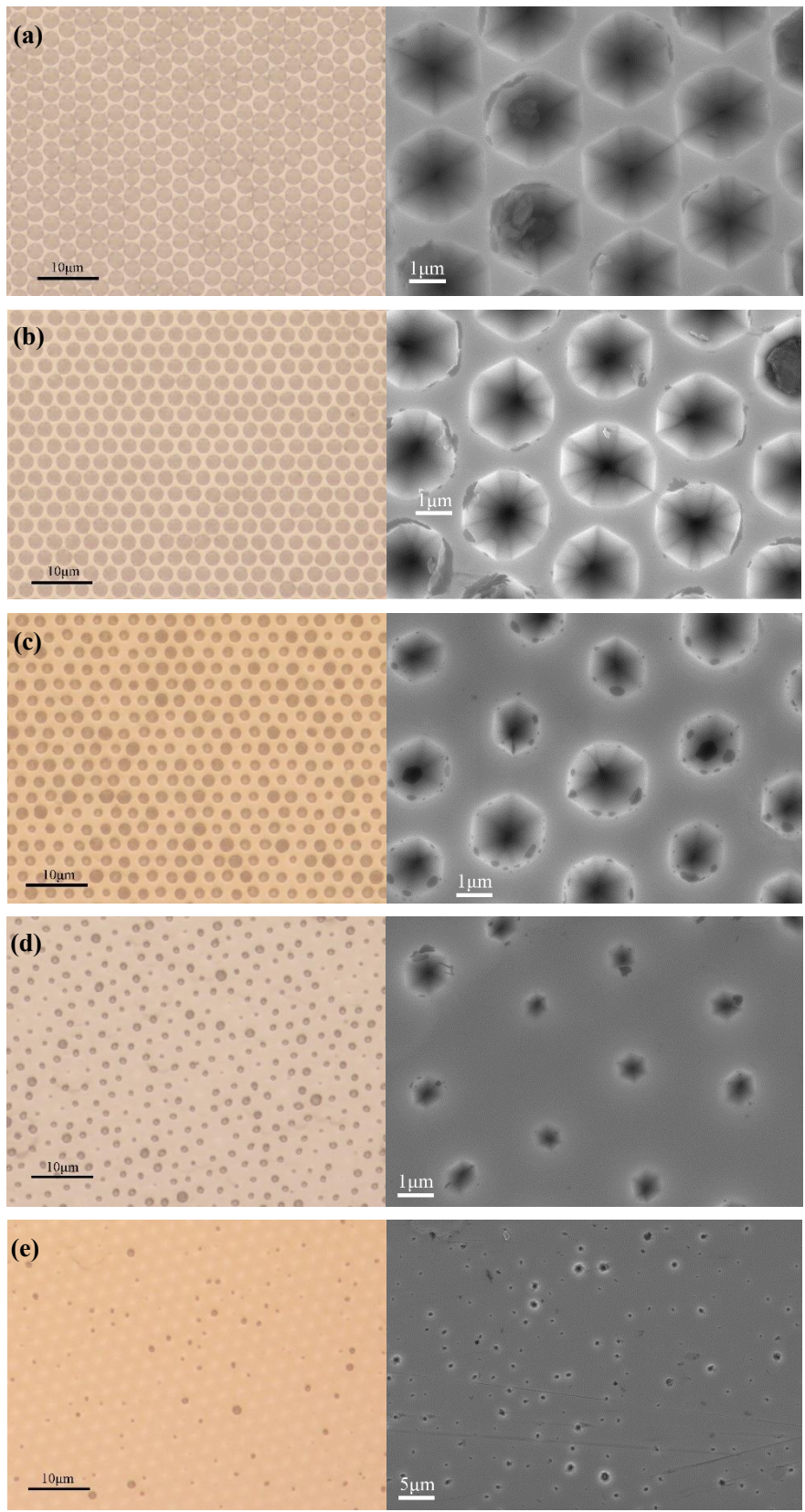

Fig.3 L. Wang *et al.*